\documentstyle[11pt,newpasp,twoside,epsf]{article}

\begin{document}

\title{Possible pulsations of the M giant in MWC560}

 \author{Sylwia M. Fr\c{a}ckowiak, Maciej Miko{\l}ajewski, 
Toma Tomov}

\affil{
Centre for Astronomy, Nicolaus Copernicus University,\\
Poland, 87100 Toru\'{n}, ul. Gagarina 11
}

\begin{abstract}
MWC 560 and other peculiar symbiotic binary, CH Cyg, have been classified
as the prototypes of the new subclass of symbiotic systems---propellers
 with accretion onto a magnetic white dwarf being the source of activity
(Miko{\l}ajewski, Miko{\l}ajewska \& Tomov, 1996). The companion of the hot component in
this model should demonstrate stellar wind with relatively high mass-loss
$10^{-5} - 10^{-7} M_{\rm Sun}/{\rm year}$. It can be expected rather a late AGB, Mira or semiregular
variable (SR) than any normal red giant.Our analysis performed for the $i$ band shows that the M giant probably belongs to
such kind of stars and pulsates with a period of about 5 months.

\end{abstract}

Unfortunately, the nature of the M giant in MWC 560 is uncertain.  No
significant variability has been observed yet. The spectral classification from
the optical is affected by the veiling blue continuum and gives probably to
high, M3-4 spectral class (e.g. Allen, 1978). In the infrared the M giant
spectrum seems to be free from the hot component, it is however affected by the
dust emission.  The spectral type estimations give rather a later M4-5 giant
(Thakar \& Wing, 1992). The mean values of the infrared indices $J-H =
1.00$ and $H-K = 0.33$ (Zhekov et al., 1996) also can indicate almost the
latest spectral types  between unreddened M6 and M5 with reasonable reddening
$E_{B-V} = 0.25$. These estimations have been very good confirmed by spectral
classification carried out using five TiO bands in the infrared (M\"urset \& Schmid, 1999).
 Their results were M5.5 and M6 obtained in two different
dates. 

We have observed MWC 560 during 1992---1999 for looking of the long-term
variability with $UBVri$ filters. The observations were carried out with 60cm
telescope at Toruñ Observatory, equipped with the one-channel diaphragm
photometer and unrefrigerated EMI 9558B photomultiplier. $UBV$ bands are very
close to the standard Johnson system whereas our instrumental $ri$ band were
significantly blueshifted until 6390A and 7420A mean (effective) wavelength
respectively. 
Our $UBV$ observations were complemented
by the large set of photometric data presented by Tomov, Kolev \& Ivanov (1996) and references therein, and 
$ri$ data by few measurements presented by Feast (1990), Buckley et al. (1990), Wing \& Landolt (1990). 

We have carried out the Fourier analysis of all data using
original Deeming (1975) and Lomb-Scargle (Lomb, 1976; Scargle, 1982) methods.
Main trend in all $UBVr$ light curves shows two pronounced maxima
around JD 2448000 in 1990 and JD 2450000 in 1995 and separated by about 2000
days. These maxima correspond to found by Doroshenko, Goanskij \& Efimov (1993) 
6-years periodicity possibly reflected an orbital motion. The other
year-scale variability has rather chaotic character. Availing of natural
 (every year) gaps in observations (caused by conjunction of MWC 560
with Sun) we have decided to remove the long-term trends in observations by
means of the one-year (in fact one-season) averages. Only for very limited $RI$
data we used one average value calculated together for 1990 and 1990/91
observations. For other seasons were used our original $\Delta r$ and $\Delta i$
data in the instrumental system where HD 59380 was taken as the comparison
star. The all obtained in such way residuals have been undergone the
frequency analysis. The resulting power spectra are showed in Fig.\,\,1.

\begin{figure}
\plotone{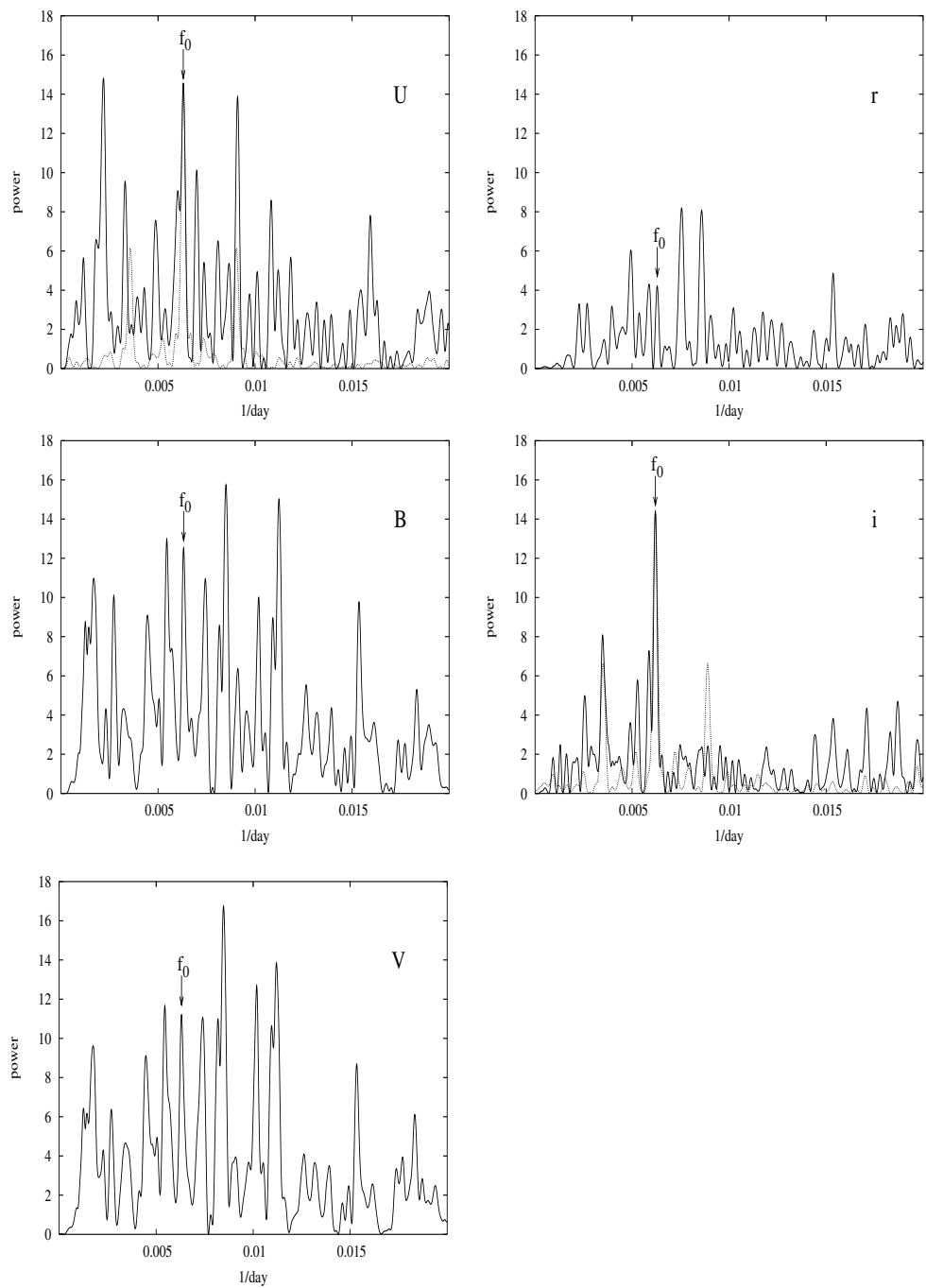}
\caption{Power spectra of residuals of $UBVri$ obserwations of MWC 560. The most interesting component ($f_0$) around 0.0062 1/day is marked by an arrow and shown for 
$U$ and $i$ light together with the spectral window (dotted line) for the $UBV$ data and $ri$ data, respectively. 
}
\end{figure}

During all period of observations (1990-99) the star varies in a very
similar way in the $UBVr$ filter with amplitudes strongly decreased from $U$ to  $r$. 
Also the periodograms (Fig.\ 1), especially for $B$ and $V$ light, are
dominated be the presence of practically the same 7 components between $0.005$
and $0.015\; 1/{\rm day}$. Most of these frequencies can be identify in $U$ power
spectrum, however they dramatically decrease or vanish in r. This behaviour
indicates that the $UBVr$ brightnesses are dominated by the hot, active
component which underwent any chaotic changes with time-scale of 6-2 months.
Only the $i$ light curve and they power spectrum significantly differ from
UBVr light curves and their periodograms. It can be understand if the
flux in our $i$ band is dominated mainly by the M giant.

Against to $UBVr$ the $i$ light power spectrum is dominated by one component
around $0.062\; 1/{\rm day}$. Its one-year aliases are also clearly visible (Fig.\ 1).
This most pronounced peak corresponds to the period $P=161.3\, {\rm days}$. It is very
probably that there is the real periodicity belonging to the M giant in this
system. 

\begin{figure}
\plottwo{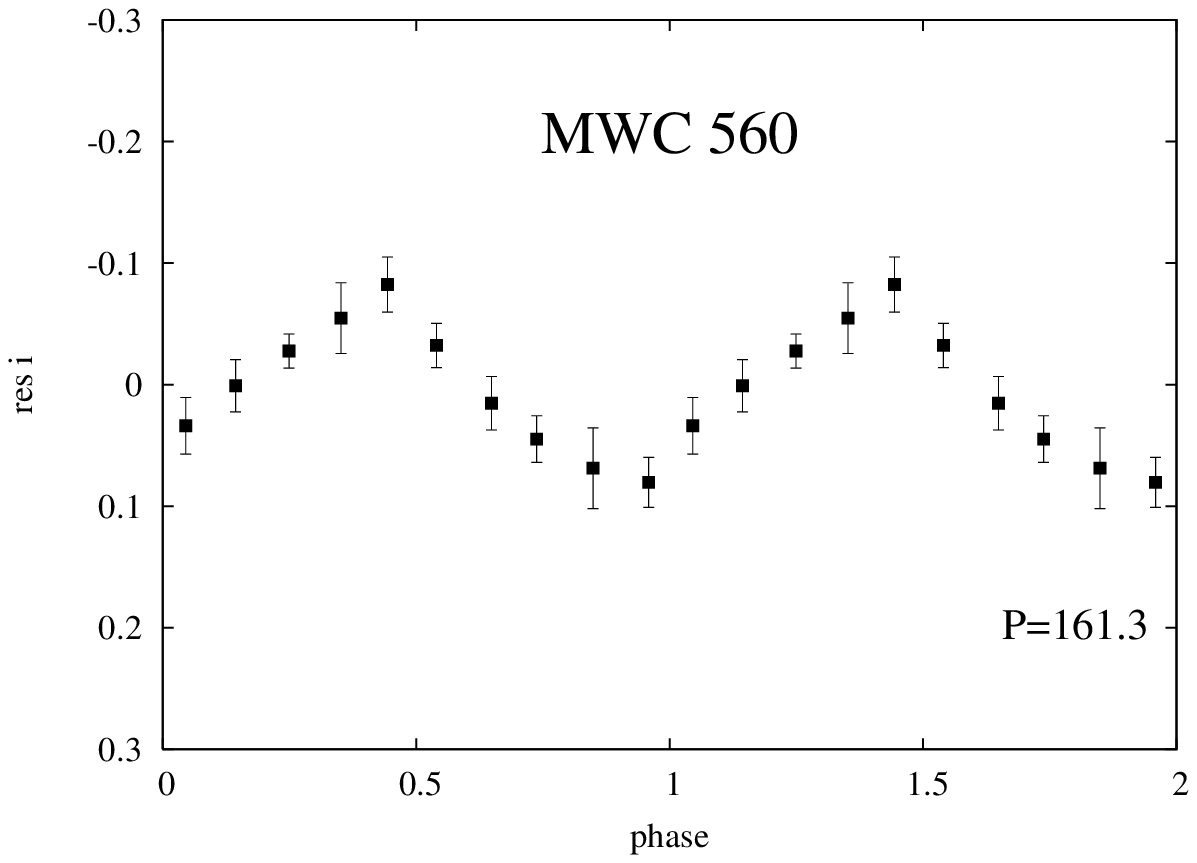}{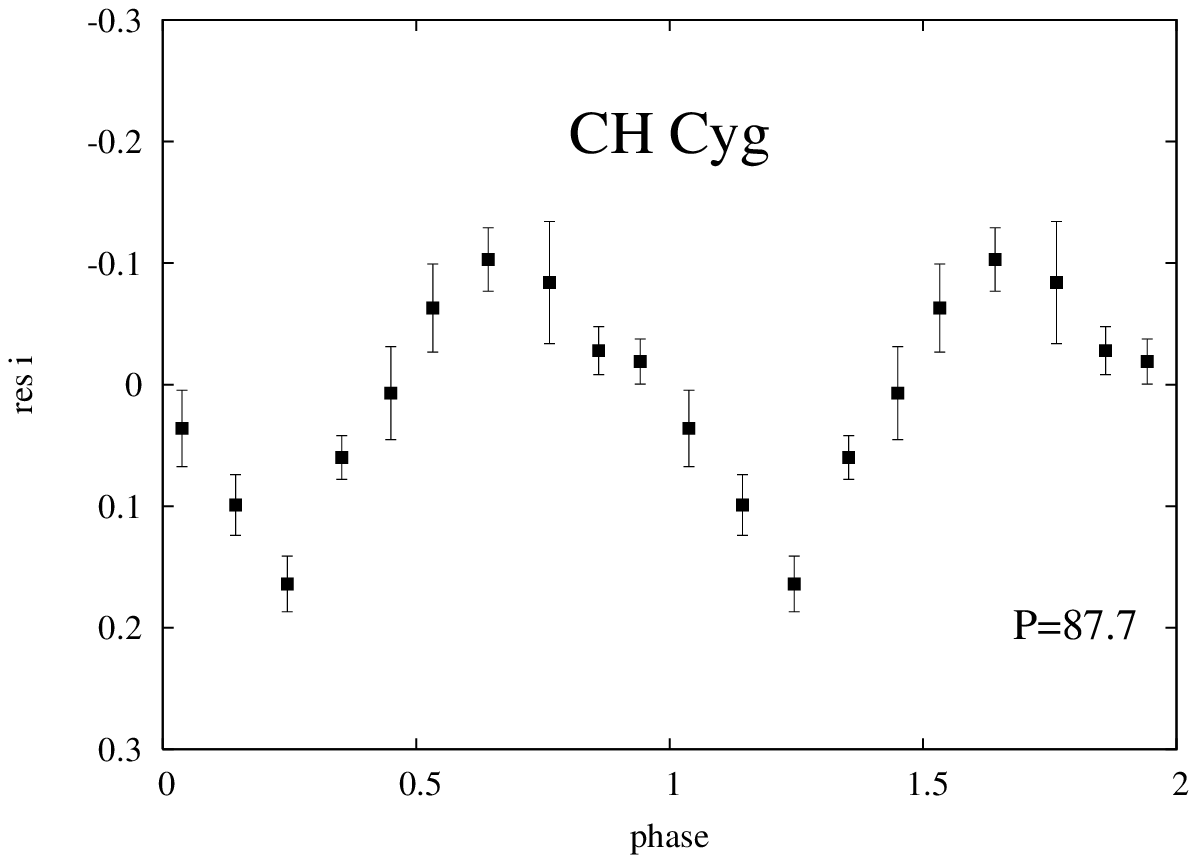}

\caption{{\it Left:} The mean light curve of MWC 560 in $\Delta i$ band folded with period
161.3 days and binned over  0.1 phases. Every points represents
4-15 individual observations.
{\it Right:} The mean light curve of CH Cyg in Toru\'{n} instrumental $i$ band folded
with the period 87.7 days during periods:
JD 2448887-2449432 and JD 2449722-2450121. The data was binned in the same
way as in MWC 560. Every point represents 4-13 individual observations.
}
\end{figure}

In Fig.\ 2 we present the mean $i$ light curve of MWC 560 binned with step of
0.1 phase of the 161.3 days period. Half-amplitude of this variability is
about 0.08 mag i.e. on about 3 sigma level.

Another twin for MWC 560 symbiotic binary seems to be CH Cyg which has been
discovered in 1924 and observed four decades as the semiregular (SR) M giant
variable with pulsation period about 100 days Miko{\l}ajewski, Miko{\l}ajewska \& Khudyakova (1990) and references therein. 
In the $V$ light this variations have average amplitude about
0.7 magnitude (Miko{\l}ajewski, Miko{\l}ajewska \& Khudyakova, 1992). For the comparison to MWC we have
found in our own observations of CH Cyg two periods JD 2448887-2449432 and
JD 2449722-24450121 when pulsations were clearly visible in $i$ band. On the periodogram we have
found the best period for this time: $P=87.7\, {\rm days}$. The mean $i$ light curve of
CH Cyg is shown in the right panel on Fig.\ 2. It is clear that both stars
seem to pulsate in similar way. The lower amplitude in MWC 560 can be caused
by additional veiling hot continuum. Such pulsations are characteristic for
semiregular pulsating red giants similar to Miras but having significantly
lesser amplitude (known as SRa stars in GCVS catalog). They are probably
relatively young AGB stars with large mass loss.

It is very interesting  that the same little shifted frequency-component
also exists on all power spectra from $U$ to $r$ light. Moreover there is 
only one such component and its amplitude seems to be systematically
decreasing from $U$ to $r$ (marked by arrow in Fig.\ 1). The $UBV$ fluxes in MWC 560 are dominated by a
hot continuum which also significantly contributed in our instrumental
blueshifted $r$ bands. It is possible that in 159 days periodicity in
hot continuum reflects variable accretion rate from the stellar wind
modulated by the donor pulsations.

\acknowledgements{ 
S.M.F. and T.T. are grateful to SOC and LOC for the support. 
This study was supported by Polish KBN
Grant No. 5 P03D 003 20.
}

\end{document}